\documentclass[amsmath,twocolumn,superscriptaddress,prx,aps]{revtex4-1} 
\usepackage{amsthm,amsfonts,graphicx,verbatim, color}
\usepackage[singlelinecheck=false]{caption}
\usepackage{bm}
\usepackage[utf8]{inputenc}
\usepackage{ulem}
\let\oldmarginpar\marginpar
\renewcommand\marginpar[1]{\-\oldmarginpar[\raggedleft\footnotesize #1]%
{\raggedright\footnotesize #1}}

\newcommand{\be}{\begin{equation}}
\newcommand{\ee}{\end{equation}}
\newcommand{\bea}{\begin{eqnarray}}
\newcommand{\eea}{\end{eqnarray}}

\renewcommand{\epsilon}{\varepsilon}

\def\beq{\begin{equation}}
\def\eeq{\end{equation}}
\def\bea{\begin{eqnarray}}
\def\eea{\end{eqnarray}}

\begin{document}

\title{Lifetimes of local excitations in disordered dipolar quantum systems}
\author{Rahul Nandkishore}
\affiliation{Department of Physics and Center for Theory of Quantum Matter, University of Colorado at Boulder, Boulder CO 80309, USA}
\author{Sarang Gopalakrishnan}
\affiliation{Department of Physics, Pennsylvania State University, University Park PA 16802, USA}

\begin{abstract}

When a strongly disordered system of interacting quantum dipoles is locally excited, the excitation relaxes on some (potentially very long) timescale. We analyze this relaxation process, both for electron glasses with strong Coulomb interactions---in which particle-hole dipoles are emergent excitations---and for systems (e.g., quantum magnets or ultracold dipolar molecules) made up of microscopic dipoles. 
We consider both energy relaxation rates ($T_1$ times) and dephasing rates ($T_2$ times), and their dependence on frequency, temperature, and polarization. Systems in both two and three dimensions are considered, along with the dimensional crossover in quasi-two dimensional geometries. A rich set of scaling laws is found. 
\end{abstract}
\maketitle

\section{Introduction}

The dynamical behavior of well isolated and strongly disordered quantum many body systems has been arousing great interest, from both the condensed matter and quantum information communities. Under certain circumstances, such systems can display the phenomenon of `many body localization' (MBL) \cite{Anderson, fleishman, GMP, BAA} (for reviews, see \cite{MBLARCMP, mblrmp, gpreview}), whereby they preserve a memory of their initial conditions forever in local observables, hence serving as good quantum memories. The existing theory of MBL describes systems with interactions that are {\it short range} in real space, whereas physical systems frequently contain {\it long range} interactions---e.g., Coulomb or dipolar interactions. 
For interactions that fall off as a power law with distance, it has been argued that MBL is always ultimately destabilized by nonperturbative processes due to rare regions \cite{avalanche, ldrh}. However, such putative non-perturbative effects only manifest on timescales which diverge faster than a power law of the disorder strength~\cite{gh2019}, and which as such are not relevant for most experiments in the strongly disordered regime (but see Ref.~\cite{leonard2020signatures}). We will not consider such non-perturbative effects in this paper. However, 
for sufficiently slowly decaying power laws, there is also a perturbative instability of the MBL phase \cite{PhysRevLett.80.2945, Burin, yaodipoles, PhysRevB.91.094202, PhysRevB.97.214205, PhysRevB.101.024201} (although see \cite{LRMBL} for exceptions), which {\it is} expected to manifest on experimentally relevant timescales. 
Even when the MBL phase is perturbatively unstable on the longest timescales, however, the framework of ``nearly MBL'' systems offers a useful perspective on the dynamics of well isolated but strongly disordered long range interacting systems---which remains an important open problem. 

{\it Dipolar} systems present a particularly interesting subclass of long range interacting systems. Strongly disordered systems of interacting dipoles arise \textit{microscopically} in multiple experimentally relevant contexts, including: 
 (i)~dipolar molecules in optical lattices~\cite{yan2013observation};
 (ii)~dense ensembles of nitrogen-vacancy centers in diamond~\cite{kucsko1, kucsko2, Lukin};
 (iii)~dipolar quantum magnets, such as lithium holmium fluoride~\cite{ghosh2002coherent}. In addition, \textit{emergent} dipolar excitations dominate the low energy physics of the electron glass \cite{electronglassreview}. Dipoles at non-zero temperature in two and three dimensions are known to exhibit a perturbative instability to MBL \cite{Burin, yaodipoles}. At the same time, the presence of a long range (and highly anisotropic) interaction makes dipolar systems challenging for exact theoretical analysis, particularly in an out-of-equilibrium setting. However, in the limit of strong disorder, the dipoles that contribute to slow dynamics and relaxation interact via sparse, long-range resonances; in this limit, therefore, one can construct controlled theories of response and  dynamics.

The excitations of a system of interacting dipoles are in general complicated delocalized modes. However, in the strongly disordered regime, the dynamics of each dipolar excitation can be separated into ``fast'' local precession and ``slow'' relaxation through coupling to other dipoles. 
In this sense, each dipole can be regarded as a long-lived excitation of the system,
and interactions between dipoles matter because they cause individual dipoles to dephase and relax. 
The relaxational dynamics of dipolar systems can be usefully described in terms of the population relaxation time (bit flip time) $T_1$ and the dephasing time $T_2$, which in turn can be directly measured experimentally by interferometric methods~\cite{ngh, PhysRevLett.113.147204, chiaro2019direct} or via `pulsed' experiments such as 2D coherent spectroscopy \cite{HammZanni, Mukamel}. This latter experimental technique was recently applied to study the relaxational dynamics of the emergent dipolar excitations in phosphorous doped silicon \cite{electronglasspaper}, where the frequency, temperature and doping dependence of the $T_1$ and $T_2$ times was experimentally measured and explained in terms of a dipolar hopping model. Ref.~\cite{electronglasspaper} operated entirely with a three dimensional system, at effectively zero temperature (i.e. temperatures lower than experimental probe frequency). 

In this paper, we analyze the relaxational dynamics of dipolar systems across a wide range of settings. In particular, we determine the frequency, temperature, and polarization dependence of the $T_1$ and $T_2$ times for systems in two and three dimensions, as well as the dimensional crossover in quasi-two dimensional systems (slab geometries). We find that relaxation in dipolar systems, especially in two dimensions, is due to subtle collective effects, which nevertheless give rise to simple scaling laws (Table~\ref{scaletab}).  

We emphasize at the outset that frequency dependent relaxation rates are {\it not} the same as ac conductivity.
The ac conductivity describes the response of some reference equilibrium state to a weak time-dependent perturbation; at finite frequencies, it is nonzero both for metals and Anderson (or many-body) localized insulators. Even though there is no relaxational dynamics in insulators, the linear-response ac conductivity remains nonzero because it is dominated by sharp absorption resonances. Instead, relaxational dynamics has to do with the lifetimes of certain \emph{excited} states that are locally out of equilibrium: i.e., it is the timescale on which such states, once created, return to equilibrium. Formally, relaxation rates are a property of multiple-time correlation functions of spatially local operators, such as the local charge density operator or the local fermion creation operator. 
The most basic observable that measures ``relaxation'' is the linewidth of a local spectral function; however, this is often dominated by inhomogeneous broadening~\cite{electronglasspaper}, so one needs more delicate spatially local observables ~\cite{PhysRevLett.113.147204, chiaro2019direct} or pulsed techniques \cite{electronglasspaper} to identify relaxation.
The contrast between these concepts is particularly clean if one considers a non-interacting Anderson insulator. The ac conductivity is given by the Mott formula \cite{Mott} $\sigma(\omega) \sim \omega^2$. However, if one prepares an initial state where some low energy localized orbitals are unoccupied and some higher energy localized orbitals are occupied, then this initial state is nevertheless an eigenstate of the dynamics, and as such the relaxation rates are {\it zero} i.e. the state lives forever. Relaxation rates are, however, related to the zero-frequency limit of the conductivity, as well as the field-dependence of nonlinear response~\cite{kns, gkd}. Our goal in this paper is to determine relaxation rates in dipolar systems.

This paper is structured as follows. 
In Sec.~\ref{ddos} we characterize the various regimes of the electron glass and of microscopic dipolar ensembles via a density-of-states exponent $\gamma$. 
In the rest of the paper we present a unified theory of dipolar relaxation; the origin of the dipoles enters our analysis purely through the exponent $\gamma$. 
We address, in turn, three-dimensional systems (Sec.~\ref{3d}), two-dimensional systems (Sec.~\ref{2d}), and slab geometries (Sec.~\ref{slabs}); in each case we discuss both zero-temperature and finite-temperature behavior. 
Finally, we generalize our results to the case where polarization rather than temperature sets the entropy of the ensemble (Sec.~\ref{polarization}).
Throughout, we assume we are dealing with an {\it isolated} system, decoupled from phonons or any external environment. (For a discussion of imperfectly isolated near localized systems, see, e.g., \cite{mblbathgeneral}).

\begin{table}
\begin{tabular}{|c|c|c|c|} \hline
scale & $d$ & $T = 0$ & $T > 0$ \\ \hline
$T^{-1}_{1,2}$ & $3$ & $\begin{array}{lr} \omega & (\gamma = 1) \\ e^{-c/|\omega|^{\gamma-1}} & (\gamma > 1) \end{array}$ & $T^\gamma$ \\ \hline
$T^{-1}_1$ & $2$ & 0 & $\begin{array}{lr}T^{4\gamma} & (typical) \\ T^{3\gamma} & (resonant) \end{array}$ \\ \hline 
$T^{-1}_2$ & $2$ & 0 & $T^{3\gamma}$ \\ \hline
$(T^*_2)^{-1}$ & $2$ & 0 & $T^{3\gamma/2}$ \\\hline
\end{tabular}
\caption{Summary of main results for scaling of relaxation rates in three and two dimensions with frequency at zero temperature, and with temperature. For the density of states exponent $\gamma$ see Table~\ref{gammatab}.}
\label{scaletab}
\end{table}

\begin{table}[h]
\begin{tabular}{|c|c|c|} \hline
system & regime & $\gamma$ \\ \hline
Electron glass & Mott $\left(\frac{V}{\xi |\log\omega|} < \omega\right)$ & 2 \\
 & Shklovskii-Efros $\left(\frac{V}{\xi |\log\omega|} > \omega\right)$ & 1 \\ \hline
Dipolar spins & No continuous SSB & 5 \\ \hline
\end{tabular}
\caption{Density of states of low-energy dipolar excitations in the systems considered here.}
\label{gammatab}
\end{table}

\section{Density of states of dipoles}\label{ddos}

As we discussed in the introduction, dipolar systems arise in many physical settings. Excitation lifetimes in all of these systems are set by very similar mechanisms, but the resulting rates are sensitive to the density of states of low-frequency dipolar excitations. In all the cases of interest, $\rho(\omega) \sim \omega^{\gamma-1}$ (up to logarithmic corrections). Therefore, at low temperatures, the fraction of dipoles that are thermally active scales as $T^\gamma$. In this section we discuss the values of $\gamma$ in various cases; in subsequent sections we will present unified expressions for rates (that apply to \textit{all} the dipolar systems of interest) in terms of this exponent $\gamma$. Our results in this section are not new, but are presented for completeness.

\subsection{Electron glass}

The low energy excitations of the electron glass involve an electron moving from a low energy localized orbital to a nearby higher energy localized orbital. This localized particle-hole excitation can be thought of as an emergent electric dipole.

The reduction to dipoles is standard (see, e.g., the supplement to \cite{electronglasspaper}) but for completeness we outline the main steps here.
To fix units, consider a general microscopic Hamiltonian describing spinful electrons subject to Coulomb interactions and a random potential:
\begin{equation}\label{H0}
    H = \sum_{i\sigma} \epsilon_i n_i + \sum_{\langle ij \rangle, \sigma} t_{ij} c^\dagger_{i\sigma}c_{j\sigma} + \sum_{i \neq j, \sigma} \frac{V}{|\mathbf{r}_i - \mathbf{r}_j|} n_{i\sigma} n_{j\sigma}.
\end{equation}
The on-site energies $E_i$ are drawn from some distribution of width $W$, and the hopping terms from some distribution with characteristic scale $t$. The third energy scale in the problem is set by the Coulomb potential; we denote by $V$ the scale of the Coulomb potential at the typical nearest neighbor distance between electronic orbitals (in our units this distance is set to unity). In what follows we will essentially ignore the spin degree of freedom, which was argued to be unimportant for THz experiments in \cite{electronglasspaper}. Our focus is on relaxation of the electric dipoles. For a discussion of the spin sector, see e.g. \cite{BhattLee}. 

In the electron glass, the randomness comes microscopically from the positions $\mathbf{r}_i$. Thus, $t_{ij} = t_0 e^{-|\mathbf{r}_i - \mathbf{r}_j|/r_0}$, i.e., the wavefunction overlap between two orbitals. Since this is exponentially decaying we truncate it at the nearest-neighbor scale. Similarly, the on-site potential at each site comes from treating Coulomb interactions at the Hartree level. Since the hopping is exponentially sensitive to the inter-site spacing, while the Coulomb interaction falls off as a power law, one can increase the ratio of hopping to interactions by increasing the density of carriers. At some critical density, the single-particle states delocalize and the system undergoes a metal-insulator transition. Our analysis is confined to the insulating side of this transition. We have introduced the problem schematically, as the details do not affect the scaling behavior that is of primary interest here.

We first consider the single-particle problem, with interaction effects included (e.g.) at the Hartree-Fock level. Single-particle eigenstates are localized with a localization length $\xi$. In three dimensions this in turn requires $t<W$. There is a corresponding energy scale $\delta_\xi$, which is the level spacing between single-particle orbitals in the same localization volume. (At large $\xi$ it scales roughly as $t/\xi^3$. Deep in the localized phase, $\xi \ll 1$ so this characteristic scale saturates to $\delta_\xi \sim t$.) We will only consider temperatures and frequencies $\omega, T \ll \delta_\xi$. At these low frequencies, there are no excitations in a typical window of the sample, i.e., it looks locally ``gapped.'' The excitations that do exist in this frequency window are those involving charge transfer over a scale that is large compared with $\xi$. Such large-scale rearrangements can be arbitrarily low-energy; however, a local operator has vanishing matrix elements between spatially separated states. The physically relevant low-energy excitations---i.e., those that can be excited by local probes---instead involve atypical orbitals that are centered around pairs of sites $i$ and $j$ that are much farther than $\xi$ apart, but energetically ``resonant,'' in the sense that $\epsilon_i \approx \epsilon_j$. The single-particle eigenstates on a resonant pair are (approximately) symmetric and antisymmetric linear combinations of orbitals centered at $i$ and $j$. When one of these eigenstates is occupied and the other is empty, the resonant pair is a two-level system with a dipole moment set by its radius $r = |\mathbf{r}_i - \mathbf{r}_j|$. A well-known result due to Mott is that resonant pairs, or dipoles, with a transition frequency $\omega$ have a typical size $r_{\omega} \sim \xi \log(W/\omega)$. The effective Hamiltonian for these emergent ``active'' dipoles takes the form 
\begin{equation}\label{cpair}
    H_{\mathrm{pair}} = \sum_\alpha \mathcal{E}_\alpha \tau^z_\alpha + \sum_{\alpha\neq \beta} \frac{ c'V p_\alpha p_\beta}{r_{\alpha\beta}^3} (\tau^+_\alpha \tau^-_\beta + \mathrm{h.c.}). 
\end{equation}
where $ c'$ is an prefactor of order unity that is not important for the argument, the $\tau$ operators are Pauli operators acting on the emergent dipoles, $\alpha$ labels emergent dipoles, and $p_\alpha \sim \xi\log(W/\mathcal{E}_\alpha)$ is the size of emergent dipole $\alpha$. 

It remains to establish the density of states of these active dipoles. There are two low-frequency regimes:

\begin{description}

\item[Mott regime] When the transition frequency $\omega$ is relatively large, i.e., $\omega \agt V/|r_\omega|$, interactions do not affect the occupation numbers of the relevant single-particle eigenstates. For a resonant pair to be active, the lower eigenstate of the pair must be within $\omega$ of the Fermi energy. The density of resonant pairs at frequency $\omega$ then scales as $\omega n \xi r_\omega^{d-1}/W^2$, where $r_{\omega} \sim \xi \log (W/\omega)$ and $n$ is the dipole density.

\item[Shklovskii-Efros regime] When $\omega \alt V/|r_\omega|$, interactions qualitatively rearrange the ground state through a mechanism similar to Coulomb blockade. As long as \emph{either} site forming a resonant pair has a single-particle energy within $V/|r_\omega|$ of the Fermi energy, the doubly occupied state is energetically unfavorable, so the pair of sites is singly occupied in the ground state. Therefore, the phase space for the TLS goes as $Vn \xi| r_\omega|^{d-2}/W^2$, which is essentially constant at low frequencies~\cite{shklovskii1981phononless}. 

\end{description}

In addition to these ``low-frequency'' regimes where localized dipolar excitations are well-defined, there is a high-frequency regime where the eigenstates of~\eqref{cpair} are delocalized and local dipolar excitations relax rapidly. We will not consider this high frequency regime. 
Note that the Shklovskii-Efros regime always exists and describes the low energy density of states of the electron glass. The `Mott' regime exists IFF the crossover frequency set by the relation $V/r_{\omega} = \omega $ is below the upper cutoff $\delta_{\xi}$ i.e. IFF $\frac{V}{\xi \log(W\xi/V)} < \delta_{\xi}$.

To summarize, in the electron glass we generically have $\gamma = 1$ at the lowest frequencies, but there is potentially also an intermediate-frequency ``Mott'' regime for which $\gamma = 2$.

\subsection{Microscopic dipoles}

We now turn to systems consisting of microscopic dipoles, such as quantum magnets with strong dipolar interactions~\cite{aeppli}, lattices of ultracold dipolar molecules, and dense ensembles of nitrogen-vacancy centers in diamond. 
%
%
%
The $T = 0$ behavior of excitations in such random spin systems can be classified according to whether they break a continuous symmetry or not~\cite{GurarieChalker}; in either case, the excitations are generically bosonic in character, and can be thought of as spin waves in a random medium.

If the ground state spontaneously breaks a continuous symmetry, the low energy excitations are Goldstone modes. At nonzero frequency the Goldstone modes would be localizable, absent the dipolar interaction; the long-range tail of the dipolar interaction creates resonances among these localized orbitals. (In fact, even short-range interactions are expected to cause delocalization in this limit, since the nearly delocalized Goldstone modes can act as a bath~\cite{NandkishorePotter}.) Thus, there are no well-defined \emph{localized} excitations at low frequency in this case, and we will not consider it further.
In the case where the ground state does not spontaneously break a continuous symmetry,  there are no Goldstone modes, and all single particle localization lengths are bounded at strong disorder. In this case it universally follows \cite{GurarieChalker} from stability of the ground state that the low energy density of states for non-Goldstone bosonic excitations in random media must scale as $\rho(\tilde \omega) \sim \tilde \omega^4$. Thus for generic systems of microscopic dipoles that do not spontaneously break a continuous symmetry, we have $\gamma = 5$. 

\section{Relaxation in three dimensions}\label{3d}

In the rest of this paper we will take the dipolar density-of-states exponent $\gamma$ as an input, and construct a general theory of dipolar relaxation. In three dimensions, the long-range tail of the dipolar interaction guarantees that all excitations will eventually delocalize~\cite{LevitovDipoles}, but the timescales diverge at low frequencies and temperatures. 

\subsection{Zero-temperature relaxation}

We first consider zero-temperature relaxation. (More generally this section concerns relaxation of excitations above \emph{zero-entropy states}, which could also, e.g., be fully polarized states that are far from the ground state.) This occurs through the long-range hopping of a single dipolar excitation of frequency $\omega$. 
Within perturbation theory in the dipolar interaction, an excitation can only hop to another of similar frequency---for example, an excitation of frequency $\omega$ can only hop to other states in the frequency window $(0, 2\omega)$, which occur with density $\sim \omega^\gamma$. The distance between neighboring dipolar excitations therefore scales as $\omega^{-\gamma/3}$, and the interaction between them scales as $\omega^\gamma$. If $\gamma = 1$, the typical interaction between excitations at frequency $\omega$ scales the same way as the typical detuning (this case is therefore ``marginal''~\cite{electronglasspaper}). If $\gamma > 1$, on the other hand, neighboring low-frequency excitations are detuned by much more than the interaction between them. The dynamics in these two cases is therefore very different, and we consider them in turn. 

If $\gamma > 1$, one can estimate the characteristic lifetime as follows. Starting from a typical spin the number of resonances within a radius $R$ scales as $\log R$~\cite{LevitovDipoles}. 
The criterion for finding a resonance is that the number of partners times the matrix element should equal the typical detuning, i.e., $\omega^\gamma \log R \sim \omega$, which gives
\begin{equation}
\log R \sim \omega^{1-\gamma} \Rightarrow \Gamma_\omega \sim \exp(-c |\omega|^{1-\gamma}).
\end{equation}
Thus the excitation lifetimes diverge with an essential singularity at low frequencies. 

If $\gamma = 1$, on the other hand, the detuning and matrix element scale the same way with distance. Thus a dipole at low frequency $\omega$ finds a resonant partner at a distance of order $\omega^{-1/3}$---i.e., it only depends on $\omega$ through the overall density of the resonant network. Thus an excitation decays at a rate
 $\Gamma \sim \omega$: %
a scaling that was described as the ``marginal Fermi glass''~\cite{electronglasspaper}. 
In this marginal case, logarithmic corrections play an important part at low frequencies. In the specific case of the electron glass at low temperature, there are logarithmic enhancements to the density of states for dipoles as well as the dipole moments (and thus the interaction between dipoles), as discussed in the previous section. Putting these together we find the scaling
\begin{equation}\label{mgscale}
\Gamma(\omega) \sim \omega |\log^3 \omega|.
\end{equation}
We discuss the numerical prefactors in the Appendix. Note that Eq.~\eqref{mgscale} implies that at very low frequencies, individual particle-hole excitations are not well-defined but instead become strongly coupled. Thus (within our analysis) the ultra-low-temperature state is apparently a ``non-Fermi-glass'' whose excitations are highly collective in terms of the original fermions (though they are simple if one regards them as delocalized dipoles). However, this regime does not onset until exponentially low temperatures and frequencies, which may be difficult to access experimentally. 

\subsection{Finite-temperature relaxation}

At finite temperature, interactions between dipoles are a crucial additional channel for relaxation. 
The effective Hamiltonian for interacting dipoles takes the form \cite{yaodipoles}
\begin{equation}\label{eq:Burin}
    H_{\mathrm{pair}} = \sum_\alpha \mathcal{E}_\alpha \tau^z_\alpha + \sum_{\alpha\neq \beta} \frac{J}{r_{\alpha\beta}^3} (\tau^+_\alpha \tau^-_\beta + \mathrm{h.c.} + \lambda \tau^z_{\alpha} \tau^z_{\beta}), 
\end{equation}
which is the same up to relabellings as Eq.\ref{cpair} except for a $ZZ$ interaction (which can be ignored at zero temperature, when there is only a single excited dipole). 

At finite temperature, 3D dipolar systems host a bath of delocalized thermal excitations, as identified by Burin~\cite{Burin}. We present this argument for general dimensions $d$, then specialize to $d = 3$. In general, the density of dipolar excitations of size $R$ in a $d$-dimensional sample with dipolar interactions scales as $R^{d - 3}$. At temperature $T$, the density of thermally excited dipoles scales as $T^\gamma R^{d-3}$. Since all of these excitations have similar detunings, they hybridize if they overlap. If we pick one resonance, the number of thermally excited overlapping excitations scales as $T^\gamma R^{2d - 3}$. There is an associated thermal length $R(T) \sim T^{-\gamma/(2d-3)}$ at which thermally excited resonances form a percolating hopping network with characteristic timescale
\begin{equation}\label{burintime}
\tau_{\mathrm{Burin}} \sim T^{-3\gamma/(2d-3)}.
\end{equation}
Specializing to three dimensions, we find that the characteristic timescale of the Burin resonances simply scales as $T^{-\gamma}$. 

In three dimensions, the density of the percolating network (as computed above) coincides with that of thermally activated spins. Thus, for scaling purposes, one can regard all thermally active spins as being on the resonant network. (We will see below that the situation in two dimensions is drastically different.) We now consider the relaxation of an inserted excitation at frequency $\omega$. To relax, this excitation must find a partner that is within $T^\gamma$ of it in energy. The nearest such excitation is at $R \sim T^{-\gamma/3}$. A Golden Rule calculation yields the result that $\Gamma \sim T^{\gamma}$ from this mechanism. 

To understand the low-temperature limit, one should also address how the thermal bath interferes with coherent zero-temperature hopping. Experimentally~\cite{electronglasspaper}, going to finite temperature seems to suppress coherent hopping, as one might expect~\cite{leggett_review}. Thus excitation lifetimes are non-monotonic. Although this suppression can be quantitatively significant, it is not expected to be \emph{parametric} and we will not address it here. At our level of analysis, the excitation hops to the nearest zero-temperature or bath-induced resonance, whichever is closer. So we conclude that 
\begin{equation}
\Gamma \sim 
\left\{ \begin{array}{lr} 
\min\left[\omega, T\right] & \gamma = 1, \\
\min\left[\exp(-c/|\omega|^{\gamma-1}), T^\gamma\right] & \gamma > 1. \end{array} \right.
\end{equation}
Thus in the low-temperature Shklovskii-Efros regime of the electron glass, the effects of finite temperature are relatively obvious (viz. we just replace frequency by temperature), but away from this regime low temperatures can dramatically enhance the energy relaxation rate.

We now comment briefly on dephasing times. In the three-dimensional case, the interaction of a typical degree of freedom with the nearest thermal excitation scales as $T^\gamma$. The fluctuations of thermal excitations cause dephasing, so in the present case dephasing scales parametrically the same way as energy relaxation, so that $T_2 \simeq T_1$. (At zero temperature there is no separate dephasing channel, so $T_2$ saturates the bound $T_2 \leq 2 T_1$.

\section{Relaxation in two dimensions}\label{2d}

The dipolar interaction in two dimensions falls off sufficiently fast that single-particle excitations can be localized. In the orthogonal symmetry class (relevant for most systems of dipoles), localization occurs even for weak disorder. For the strongly disordered systems we are considering here, single dipolar excitations can be taken to be tightly localized, so that interactions between thermally excited dipoles are crucial in determining excitation lifetimes. Since zero-temperature relaxation is absent, we turn directly to the case of finite temperatures. 

In two dimensions, the Burin mechanism outlined above~\eqref{burintime} leads to the conclusion that the spacing of the percolating cluster $R(T) \sim 1/T^{\gamma}$, so that the characteristic timescale~\eqref{burintime} of the ``bath'' formed by the percolating network is 
\begin{equation}
\tau^{\mathrm{2D}}_{\mathrm{Burin}} \sim T^{-3\gamma}.
\end{equation}
Note a crucial difference from the three-dimensional case: while the density of thermal spins scales as $T^\gamma$, the spins on the percolating network are parametrically sparser, with their density scaling as $T^{2\gamma}$. It is therefore important to distinguish between three types of excitations: (i)~the inserted excitation at frequency $\omega$ whose lifetime we are considering, (ii)~generic thermal excitations, which are not part of the percolating network, and (iii)~excitations on the percolating network. While excitations on the percolating network have a lifetime $\sim T^{-3\gamma}$, ``off-network'' excitations relax much more slowly, as we will now discuss.

One can make a naive estimate of the lifetimes of off-resonant spins along the following lines. The interaction between a typical spin and the nearest network spin scales as $T^{3\gamma}$, and so does the maximum energy exchange allowed by the bandwidth of the network. To find an allowed transition, the typical spin must find a partner detuned from it by $\sim T^{3 \gamma}$. The nearest such partner is at distance $R \sim T^{-3\gamma/2}$, giving the dipolar matrix element $T^{9\gamma/2}$ and the Fermi's Golden Rule rate $\Gamma \sim T^{6\gamma}$. While this is certainly a \emph{lower bound} on the relaxation rate, we argue that relaxation in fact occurs much faster, because the estimate above neglects \textit{spectral diffusion}. 

Spectral diffusion~\cite{Galperin, gornyi2017spectral, PhysRevB.97.214205} is a phenomenon by which the frequency of a spectral transition shifts over time because the relevant degree of freedom is interacting with other, slowly fluctuating, degrees of freedom. In the system we are considering, thermally active spins are present at density $T^\gamma$, and interact with each other via static shifts of typical size $T^{3\gamma/2}$. Thus, a spectral line averaged over a very long time would have an apparent linewidth of $T^{3\gamma/2}$. This apparent broadening, conventionally denoted $1/T_2^*$, can in principle be undone using spin echo. It is therefore not a true decay timescale: the true decay timescales are $1/T_2$ (the timescale on which the spectral line shifts its frequency) and $1/T_1$ (the timescale on which the excitation actually decays). When $T_2 \gg T_2^*$, the apparent width of the spectral line interpolates from $1/T_2$ (for short averaging times) to $1/T_2^*$ (for long averaging times). 

We estimate $T_1$ in this case using two distinct approaches, which agree on the final answer. First, we note that the level spacing at the Burin scale $R \sim 1/T^\gamma$ goes as $1/R^2 \sim T^{2\gamma}$; meanwhile, the broadening of each level due to spectral diffusion scales as $T^{3\gamma/2}$, which is parametrically larger than the level spacing. Therefore, spectral diffusion randomly brings each individual dipole into and out of resonance. Assuming spectral diffusion is effective, each level is part of the ``network'' about $T^\gamma$ of the time. It relaxes at the rate $T^{3\gamma}$ while it is on resonance, thus giving a total relaxation rate $T^{4\gamma}$. 

We can also arrive at this result using a more direct self-consistent approach, as follows. We are concerned with the problem of a two-level system (consisting of the ``source'' and ``destination'' sites of the dipolar hopping pair of interest) coupled to a bath of thermal fluctuators (i.e., thermally active spins). Thus its Hamiltonian is of the form $H = h(t) \sigma^z + \Delta \sigma^x$, where $h(t)$ is the time-dependent field due to thermal fluctuators. These fluctuators come in two flavors: (i)~``network'' spins that fluctuate on a timescale $T^{3\gamma}$ and shift the energy of the level by a corresponding amount $T^{3\gamma}$, and (ii)~``off-network spins'' that shift the energy by a much larger amount $T^{3\gamma/2}$ but also fluctuate much more slowly at some as-yet-unspecified rate $1/T_2$ (which could potentially be as slow as $T^{6\gamma}$, from the naive estimate above). 
The power spectrum of the noise coming from type~(ii) spins has the form $\langle h(\omega, t)^2 \rangle \sim f(\omega - \omega_0(t))$, where $f$ is a function peaked at $\omega_0(t)$ and of width $\sim 1/T_2$, and $t$ is a much longer timescale (associated with spectral diffusion), such that $\langle \omega_0(t) \omega_0(0) \rangle \sim (1/T_2^*) \exp(-t/T_2)$. 
For a given power spectrum of $h(t)$ this two-level system has a decay rate $1/T_1$. However the power spectrum itself is implicitly a function of $1/T_1$ (which is also the rate at which the thermal fluctuators flip), while the off diagonal matrix element $\Delta$ is implicitly a function of the power spectrum (since the bandwidth of the noise governs how far in space the nearest accessible partner is). 

Since the decay rate and spectral lineshape depend nontrivially on each other, we solve for both of them self-consistently. In principle one can do this explicitly for the two-level system specified above, following Refs.~\cite{PhysRevE.79.050105, PhysRevLett.119.046601}. To simplify our analysis we assume that the noise spectral function can be characterized by a time-dependent apparent bandwidth $\Gamma(t)$ that interpolates between $1/T_2$ and $1/T_2^*$ by ``filling in'' a new spectral strip of width $1/T_2$ at a rate $1/T_2$. Thus
\begin{equation}\label{gammat}
\Gamma(t) = \left\{ \begin{array}{lr} 
1/T_2 & t \simeq T_2 \\
t/T_2^2 \quad & T_2 \ll t \ll T_2^2/T_2^* \\
1/T_2^* & t \gg T_2^2/T_2^* \end{array} \right.
\end{equation}

To set up the rest of the self-consistency loop we first estimate $T_1$ as a function of $\Gamma(t)$, which (for consistency) must itself be evaluated over the timescale $T_1$ on which a decay process happens. We estimate $T_1$ using the Golden Rule: to find a spin at detuning $\Gamma$ one must go a distance such that $R^2 \sim 1/\Gamma$, so the matrix element is $\Gamma^{3/2}$ and therefore, using the Golden Rule,
\begin{equation}\label{t1gamma}
1/T_1 \sim \Gamma(T_1)^2
\end{equation}
Finally we need to evaluate $T_2$. We make this estimate self consistently following ~\cite{ngh}. At some distance $R$ from the central spin, there are $T^\gamma R^2$ other thermally excited spins, each flipping at the rate $1/T_1$, and $T^{2\gamma} R^2$ spins {\it on the Burin network} flipping at the rate $T^{3\gamma}$.  The dipolar coupling at distance $R$ is $V/R^3$. We fix $R$ self-consistently so that $V/R^3 = (1/T_1) T^{\gamma} R^2 +  T^{3\gamma} T^{2\gamma} R^2$: once we have included the effects of thermal fluctuations to this distance, the line has broadened by enough that it is no longer able to resolve more distant fluctuations. This yields the ultimate expression  
\begin{equation}
1/T_2 \sim ( T^{\gamma}/T_1 + T^{5 \gamma})^{3/5}, \label{eq: t22d}
\end{equation}
One can check that the set of equations~(\ref{gammat},\ref{t1gamma},\ref{eq: t22d}) has the following self-consistent scaling solution:
\begin{equation}
1/T_1 \sim T^{4\gamma}, \quad 1/T_2 \sim T^{3\gamma}.
\end{equation}
This seems to be the \emph{only} self-consistent solution: if we suppose $\Gamma(T_1) \sim 1/T_2$ then the resulting decay rate is too slow for Eq.~\eqref{gammat} to be consistent, whereas if we assume $\Gamma(T_1) \sim 1/T_2^*$ the decay rate is too fast.

An interesting aspect of these results is that the effect of the Burin network spins on the decay of typical spins is \emph{marginal}, scaling the same way in Eq.~\eqref{eq: t22d} as the self-generated contribution to $T_2$. Thus the 2D system is on the cusp of being a self-sustaining metallic state through spectral diffusion. (However, absent the Burin network there would also be a self-consistent solution with all decay rates equal to zero, which the present problem cannot have.)

\section{Dimensional crossovers}\label{slabs}

We now turn to the crossover between the 2D and 3D behaviors discussed above, in a slab that has finite thickness $\ell$ in the $z$ direction, but infinite in the $x-y$ plane. 
(This can be achieved either by making a finite-width sample or by imposing a strong magnetic or electric field gradient along one direction (say the $z$ axis) of a three-dimensional sample, thus detuning dipoles at different $z$. In the latter setup the physics discussed in \cite{shattering, fractonhydro} would also come into play.) 

\subsection{Slabs at zero temperature}

At asymptotically low frequencies, a slab behaves as a two-dimensional system; in particular, in the orthogonal class, all dipolar excitations are weakly localized. However, at weak disorder, the localization length diverges exponentially in the slab width $\ell$. Thus there is a qualitative distinction between two regimes of dipolar hopping: a regime in which each dipole finds a resonant partner using the long-range three-dimensional dipolar interaction \emph{before} the crossover set by $\ell$, and which acts effectively like a metal; and a regime in which $\ell$ is shorter than the resonance scale $R_c \sim \exp(c/|\omega|^{\gamma - 1})$ [generic case] or $R_c \sim 1/\omega^{1/3}$ [marginal case], so that excitations remain tightly localized at zero temperature. (In the electron glass these crossovers all happen when the pair radius $r_\omega \ll \ell$.) 

The resonance scale $R_c$ increases monotonically as $\omega$ decreases; therefore a slab of thickness $\ell$ has a frequency $\omega^*(\ell)$ below which dipoles become manifestly sharp. For the electron glass, at large $\ell$ this crossover always happens in the Shklovskii-Efros regime, with $\omega^* \sim \ell^{-3}$ whereas at small $\ell$ the crossover can happen in the Mott regime, with $\omega^* \sim 1/\log(\ell)$. We emphasize that the change in apparent dimension is a crossover rather than a true localization transition, since technically the dipoles are always localized; however, in two dimensions the localization length in the weakly localized regime is very large and systems in this regime can be regarded as effectively metallic. We expect the crossover from weak to strong localization to happen as follows: in the weakly localized regime, the lineshape of a dipolar excitation consists of a large number of very finely spaced lines that cannot be resolved in practice. As the localization length decreases, the lines become less finely spaced and therefore resolvable. Finally, in the strongly localized limit, there is essentially just one central line left with appreciable weight. In contrast, for dipolar systems in the symplectic class, there is the possibility of a true transition, with localized behavior at small $\ell$ (effectively, stronger disorder) giving way to delocalized behavior at larger $\ell$ (effectively, weaker disorder). 

\subsection{Slabs at finite temperature}

At finite temperature, as we have discussed, Burin resonances and spectral diffusion play an important role in relaxation. 
Recall that the line broadening from spectral diffusion comes mainly from the nearest thermally active fluctuator. Whether we use the two dimensional or the three dimensional formula for spectral diffusion depends on whether the distance to the nearest fluctuator is more or less than $\ell$. Thus, the dimensional crossover for spectral diffusion happens when the density of thermal excitations $n_{ex} \approx 1/\ell^3$. In contrast the `Burin crossover' happens at $n_{ex} \approx W/(J n \ell^3)$, which is a larger value of $n_{ex}$ at strong disorder (which we have assumed). Here $W$ is the disorder scale, $J$ the hopping scale, and $n$ the density of dipoles on the lattice. Thus, we enter the `two dimensional' regime for the Burin analysis at a higher temperature than that at which we enter the `two dimensional' regime for spectral diffusion. 

In this intermediate temperature range, the coupling to the nearest thermally active neighbor scales as $T^\gamma$ (setting $T_2^*$), but the nearest ``network'' spin is much further, since the percolating network density takes its two-dimensional value $T^{3\gamma}$. The spin-echo lifetime $T_2$ crosses over at a timescale that is intermediate between these limits. The two-dimensional self-consistent analysis we developed above can be extended to this situation, but there are many cases (depending on whether each scale is three-dimensional or two-dimensional) and we will not consider these in detail. 

\section{Crossovers near maximum polarization}\label{polarization}

In experiments involving cold atoms, nitrogen-vacancy centers, etc. it is often more natural to tune the polarization than the temperature \cite{HoChoi}. Since the dipolar interaction in these systems can generically be treated in the secular approximation, it approximately conserves the total number of spin-up dipoles. Thus one can regard high-polarization ensembles as effectively being at large chemical potential $\mu$. 
The analysis above carries over with minor changes to this setup, if we replace the temperature with the density of minority-state spins, and set the density-of-states exponent $\gamma = 1$. 

There is one distinction worth noting: at low temperature, the thermal excitations are disproportionately in low-energy, well localized states, whereas for high polarization they are in random states. Therefore at large polarization there is a potential additional relaxation channel due to delocalized high-energy excitations. This does not, however, matter in the strongly disordered regime considered here, since all single-particle states are assumed to be well localized.

\section{Discussion}

We have determined the frequency and temperature dependence of relaxation rates for systems of quantum dipoles, whether microscopic or emergent (as in the electron glass). 
At zero temperature, dipoles relax via long-range coherent hopping, which is possible in three dimensions (where long-range hopping resonances proliferate) but not in two dimensions. 
In three dimensions, therefore, relaxation occurs at zero temperature. For microscopic dipoles (and the electron glass in its intermediate-temperature Mott regime), quasiparticle lifetimes grow exponentially as their frequency goes to zero, so local dipoles are asymptotically sharp excitations. 
For the electron glass at sufficiently low temperatures (i.e., in its Shklovskii-Efros regime), or for dipolar spin ensembles that are at maximum polarization rather than in their ground state, the excitation lifetimes are marginal, scaling the same way as the excitation frequency up to log corrections. 
For the electron glass specifically, these log corrections enhance dipole hopping (see Appendix), and local dipolar excitations are no longer the right variables to think about the problem at exponentially low frequencies and temperatures. (What the `right' degrees of freedom are is not clear to us, but exotic physics could conceivably be involved \cite{PotterSenthil}.)
Finite-temperature effects provide an additional transport channel.

Although energy transport is (apparently) present in the zero-temperature limit, and local particle-hole excitations relax, the dc electrical conductivity of the electron glass is zero in linear response, since moving charge across the system requires \emph{uphill} transitions, which a zero-temperature environment cannot facilitate. 
The dynamics of realistic electron glasses at very low frequencies involve subtleties that go beyond the approximations we made here, and would be interesting to revisit for a more complete model that includes spin degrees of freedom, large-scale rearrangements, etc.

In two dimensions, the situation is quite different: dipolar excitations are localized and therefore infinitely sharp at all frequencies in the zero-temperature limit, and delocalize at finite temperatures only through subtle interaction effects. We found that the $T_1$ and $T_2$ times both scale as power laws of temperature, but with distinct exponents.
``Diagonal'' interaction effects (governed by the timescale $T_2^* \sim T^{-3\gamma/2}$ are parametrically stronger than relaxation effects $T_{1,2} \agt T^{-3\gamma}$, making dipoles in two dimensions at low temperatures a promising platform for exploring dynamical signatures associated with many-body localization. 
In quasi-two dimensional systems (slab geometries) there is a crossover from three dimensional behavior at high frequency/temperature to two dimensional behavior at low frequency/temperature, with crossover scales that we have identified. 
Our results in all of these cases are collected in Table~\ref{scaletab}.

It would be interesting to go beyond the scaling theory outlined herein to also calculate (potentially logarithmic) prefactors. It would also be interesting to determine the relation between relaxation times (calculated herein) and transport coefficients. We have also identified certain regimes where the `dipolar' approach employed herein breaks down - understanding the behavior in these regimes is an important open problem. Finally, thermopower coefficients might also be interesting to explore. We leave these problems to future work. 

\begin{acknowledgments}

We acknowledge useful conversations with Peter Armitage, Mikhail Lukin, and Markus M\"uller. This material is based in part (RN) upon work supported by the Air Force Office of Scientific Research under award number FA9550-20-1-0222. S.G. acknowledges support from NSF DMR-1653271. 

\end{acknowledgments}

\appendix

\section{Crossovers in the 3D electron glass}

We now explore, in some more detail, the dynamical crossovers in the three-dimensional electron glass at zero temperature. While the scaling of the relaxation rates was addressed in the main text, here we also provide some discussion of the prefactors, which play an important part in determining the crossover timescales.

\subsubsection{Mott regime}

Let us start by considering the dynamics in the Mott regime (assuming it exists). In the language of the single-particle dipolar hopping model~\eqref{cpair}, we are concerned with the dynamics of a dipolar excitation that is initially localized on resonant pair $\alpha$ (which has frequency $\mathcal{E}_\alpha = \omega$). We have postulated weak interaction i.e. the disorder is strong enough that nearest-neighbor hops are off shell. Instead, an excitation at frequency $\omega$ can only hybridize with a partner at similar (or lower) frequency. Excitations in the frequency window $(0, 2\omega)$ occur with density $\sim \omega^2 n \xi r_\omega^2/W^2$, where $n$ is the dipole density. Thus the spacing between two nearby such excitations scales as $\omega^{-2/3}$, and the dipolar interaction matrix element between them scales as $\sim V r_\omega^2 \omega^2 n \xi r_\omega^2/W^2$. Generically, however, the detuning between these excitations is $\sim \omega$ (by construction). Asymptotically, therefore, excitations at low frequencies in the Mott regime will be off-shell with their near neighbors. 
and will be parametrically longer-lived. One can estimate the characteristic excitation lifetime as follows. Starting from a typical spin the total number of resonances nearer than a distance $R$ scales as $\log R$. An excitation at frequency $\omega$ finds a resonant partner when $\log R \sim W^2/(V r_\omega^2 \omega n \xi)$. The characteristic interaction energy at this scale, and therefore the excitation lifetime, also scale as $\Gamma_\omega \sim V \exp(-\tilde c/|\omega|)$, i.e., the excitation lifetime is nonperturbative in $\omega$.

Could there be a regime with $\omega < V r_\omega^2 \omega^2 n \xi r_\omega^2/W^2$, such that typical excitations at $\omega$ were resonant, and the relaxation rate simply followed the typical matrix element between such excitations ($\sim \omega^2$)? It is straightforward to see this is not possible. The condition above can be expressed (dropping logarithmic factors which are not large by postulate) as $\omega > \frac{W}{V n \xi^2} \frac{W}{\xi^3}$, but the first term on the right hand side is larger than one (by the weak interaction postulate), and the second term is larger than $\delta_{\xi}$ (since we need $W>t$ to have single particle states localized), so we conclude that the above condition is only satisfied for frequencies larger than $\delta_{\xi}$, but $\delta_{\xi}$ was the upper cutoff for our theory. We thus conclude that the scaling $\Gamma_\omega \sim \exp(-\tilde c/|\omega|)$ holds throughout the regime $V/r_{\omega} < \omega < \delta_{\xi}.$

\subsubsection{Shklovskii-Efros regime}

The Mott regime crosses over to the Shklovskii-Efros regime at a frequency such that $\omega_{SE} r_{\omega_{SE}} \simeq V$. In the Shklovskii-Efros regime, Coulomb interactions rearrange the ground state such that any resonant pair split by $\omega < \omega_{SE}$ with either of its partners within $V/r_\omega$ of the Fermi energy is singly occupied in the ground state. One can estimate the density of states of pairs at frequency $\omega$ as follows. The first half of the pair must be within $V/r_\omega$ of the Fermi energy, and the density of such sites scales as $V/[Wr_\omega]$. Its partner is in a shell of radius $r_\omega$ and thickness $\xi$; multiplying these, we get that the density of dipoles of transition frequency $\omega$ scales as $n(\omega) \sim V n \xi r_\omega / W^2$. 

We now consider diagonalizing the Hamiltonian~\eqref{cpair} in the Shklovskii-Efros regime for states in the frequency window $(\omega/c, c\omega)$ where $c$ is some arbitrarily chosen number of order unity. These dipoles occupy a fraction of sites $\sim \omega n(\omega)$, and have a characteristic dipole moment $r_\omega$. The interaction between adjacent dipoles in this frequency window is $V r_\omega^2 \omega n(\omega) = \omega r_\omega^3 V^2 \xi / W^2 \sim \omega |\log^3 \omega|$. 
For small enough $\omega$, therefore, the interactions among nearest-neighbor dipoles at frequency $\omega$ become strong enough that they are \emph{generically} resonant. 
Below this crossover scale, set by $\omega_c \sim \exp \big(-(W/V\xi^2)^{2/3}\big)$, dipolar excitations are strongly coupled, relative to their natural frequency. 
Perturbation theory in the dipole-dipole interaction therefore ceases to apply. 
Since the dipoles are no longer the true excitations of the system, we expect them to decay on a timescale determined by the nearest neighbor dipole-dipole interaction. 
However, the nature of the true ground state, and its thermal transport properties, are delicate questions that are outside the scope of the present work. It is also worth bearing in mind that this lowest frequency regime occurs only at {\it exponentially} low frequencies (thus, exponentially long timescales) and may be hard to see in practice because of e.g. dephasing due to coupling to phonons or an external environment. 

Before this crossover at very low frequencies, we expect to see an extensive parameter range in which the ratio of interactions to bandwidth is essentially constant. This is the regime that was recently identified as the ``marginal Fermi glass''~\cite{electronglasspaper} and seen in THz spectroscopy. In this marginal regime, a dipole finds a resonant partner on a scale $\log R \sim W^2/(V^2 r_\omega^3 \xi)$ that (up to the overall density of the network) does not depend on $\omega$. In systems without a well-developed Mott regime, $R$ is instead just the spacing between adjacent dipoles. Putting these two cases together and neglecting logarithmic corrections, one gets

\begin{equation}\label{lse}
\Gamma(\omega) \simeq \left\{ \begin{array}{lr} \omega \xi^4 (V/W)^2 & W/V\xi^2 = O(1) \\
\omega \exp(-c/\omega_{SE}) & W/V\xi^2 \gg 1. \end{array} \right.
\end{equation}
This is the characteristic ``marginal'' scaling of the relaxation rate identified in ~\cite{electronglasspaper}. 

\bibliography{LRMBL}
 \end{document}